\documentclass[fleqn,usenatbib]{mnras}

\usepackage{newtxtext,newtxmath}

\usepackage[T1]{fontenc}
\usepackage{ae,aecompl}

\usepackage{graphicx}	
\usepackage{amsmath}	
\usepackage{amssymb}	

\newcommand{\be}{\begin{equation}}
\newcommand{\ee}{\end{equation}}
\newcommand{\bary}{\begin{eqnarray}}
\newcommand{\eary}{\end{eqnarray}}

\title[VHE \texorpdfstring{$\gamma$}-ray/X-ray correlation observed in Mrk 421]{Reconcilement of VHE \texorpdfstring{$\gamma$}\,-ray/X-ray correlation studies in Mrk 421 and break-down at high fluxes}

\author[M. M. Gonz\'alez et al.]{
M. M. Gonz\'alez,$^{1}$\thanks{E-mail: magda@astro.unam.mx}
B. Patricelli,$^{2,3,4}$
N. Fraija$^{1}$
and J. A. Garc\'ia-Gonz\'alez$^{5}$
\\
$^{1}$Instituto de Astronom\'ia, Universidad Nacional Aut\'{o}noma de M\'{e}xico, Apdo. Postal 70-264, CDMX, M\'{e}xico, 04510.\\
$^{2}$Universit\`a di Pisa, I-56127 Pisa, Italy\\
$^{3}$INFN, Sezione di Pisa, I-56127 Pisa, Italy\\
$^{4}$Scuola Normale Superiore, I-56126, Pisa, Italy\\
$^{5}$Instituto de F\'isica, Universidad Nacional Aut\'{o}noma de M\'{e}xico, Apdo. Postal 70-264, CDMX, M\'{e}xico, 04510.\\
}

\date{Accepted XXX. Received YYY; in original form ZZZ}

\pubyear{2018}
\begin{document}
\label{firstpage}
\pagerange{\pageref{firstpage}--\pageref{lastpage}}
\maketitle
%
\begin{abstract}
Multi-wavelength campaigns have been carried out to study the correlation between the very high energy (VHE) $\gamma$-ray and the X-ray emissions in blazars but, no conclusive results have been achieved yet. In this paper, we add Milagro data to the existing VHE $\gamma$-ray data from HEGRA-CT1 and Whipple and test the consistency and robustness of the reported correlation between VHE $\gamma$-ray and X-ray fluxes in Mrk 421.
We found that at monthly time scale the correlation is robust, consistent between instruments and described as a linear function.  Furthermore, most of the fluxes on shorter time scales are consistent with the correlation within 3 $\sigma_A$ even, where $\sigma_A$ is an estimated intrinsic scatter. However, a break-down of the correlation becomes clearly evident at high states of activity with fluxes $\rm \gtrsim 2.5\times 10^{-10}\, cm^{-2}s^{-1}$ at energies above 400 GeV independently of the time scale, observational period or instrument, even for single flares, the X-ray and VHE $\gamma$-ray emissions lie on the correlation until the VHE $\gamma$-ray flux reaches values higher than the one mentioned above. We have interpreted our results within the one-zone synchrotron self-Compton model. We found that describing a single and unique $\gamma$-ray/X-ray correlation strongly narrows the range of possible values of the magnetic field $B$ when a constant value of the spectral index along the correlation is assumed.
\end{abstract}

\begin{keywords}
gamma rays: observations --- BL Lacertae objects: individual  (Markarian 421)
\end{keywords}



\section{Introduction}
The subclass of active galactic nuclei (AGN) that are most commonly detected at very high energies (VHE, E $>$ 100 GeV\footnote{http://tevcat.uchicago.edu/}) is blazars. Blazars  have a pair of relativistic jets formed by electrons, positron and nucleons flowing in opposite directions, with one of the jets oriented along our line of sight. Although the matter content of relativistic jets is in discussion \citep[a pure electron-positron and/or a barionic contamination jet;][]{2000ApJ...534..109S,2015APh....71....1F, 2012MNRAS.424L..26G, 2013ApJ...768...54B,2017APh....89...14F, 2018MNRAS.481.4461F}, their broadband spectral energy distribution (SED) is characterized by two broad, well-separated ``humps'':  one at low energy, peaking in radio through X-rays and one at high energy, peaking in $\gamma$-rays (see, e.g., \citealp{2009ApJ...691L..13D}). There is ample evidence that the low energy peak arises from synchrotron radiation from relativistic electrons within the jet (see, e.g., \citealp{1997ARA..35..445U}). However, the existing data do not allow for an unambiguous discrimination between leptonic (e.g. synchrotron self-Compton, SSC or external Compton, EC) and hadronic mechanisms for the high-energy peak \citep{2011ApJ...736..131A,2015APh....70...54F}. Furthermore, it is not clear which emission mechanism dominates the VHE emission of quiescent or low activity states and flares with different time scales and intensity (see e.g.  \citealp{2014ApJ...782..110A}).

The blazar Mrk\,421, located at a distance of 134Mpc \citep{2005ApJ...635..173S}, is one of the closest and brightest source in the TeV/Xray sky. Mrk\,421 was the first extragalactic source detected in the TeV $\gamma$-ray band  \citep{1992Natur.358..477P}.   Considered as an outstanding target to investigate the radiation mechanisms inside blazar jets, Mrk\,421 has been a potential candidate to investigate distinct labels of correlations between TeV $\gamma$ and X-ray fluxes during high and low activity states.  For instance, in high activity states  \cite{1995ApJ...449L..99M}, \cite{2008ApJ...677..906F}, \cite{2011ApJ...738...25A} and  \cite{2015A&A...578A..22A} reported TeV flares in coincidence with flaring activities in X-rays occurred in 1994 May, 2001 March, 2008 May and 2010 March, respectively. Taking into consideration the variability timescales in both wavelengths, they  reported strong correlations with no lags in the whole data set. In low activity states, Mrk421 was monitored in $\gamma$-rays and X-rays in 2009 between January and June \citep{Aleksic2015}  and in 2013 January - March \citep{2016ApJ...819..156B}. In 2009, \citet{Aleksic2015} reported a harder-when-brighter behavior in the X-ray flux,  measuring a positive correlation between TeV and X-ray fluxes with zero time lag. In 2013, \citet{2016ApJ...819..156B} reported a significant TeV/X-ray correlation on a timescale of about one week. In this paper, we test the consistency and robustness of the correlation between VHE $\gamma$-ray and X-ray fluxes in Mrk 421.

\section{Data sets} \label{sec:datasets}

\begin{figure*}
\begin{center}
\includegraphics[scale=0.9]{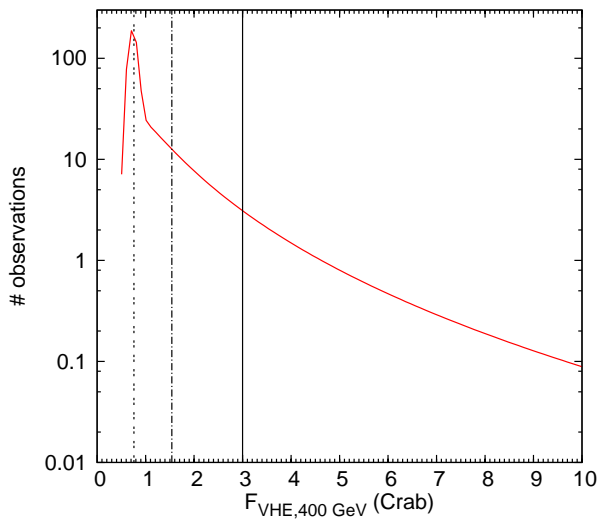}
\caption{In red: function describing the distribution of VHE $\gamma$-ray (E $>$ 1 TeV) day wise fluxes of Mrk 421 as reported by \citet{Tluczykont2010}, opportunely shifted to consider an energy threshold of 400 GeV (see text). The black dotted, dot-dashed and solid lines represent the ``very low'', ``low'' and ``high'' state flux level of Mrk 421, as measured by VERITAS \citep{2011ApJ...738...25A}.}\label{fig:Tluczykont}
\end{center}
\end{figure*}

\citet{Tluczykont2010} collected day-wise VHE $\gamma$-ray fluxes from several VHE experiments (HEGRA, H.E.S.S., MAGIC, CAT, and Whipple/VERITAS) in the time period from 1992 to 2009. From these data, a flux distribution was obtained by converting the measured flux values in units of Crab Nebula flux (as measured by each instrument) and normalizing them to a common energy threshold of 1 TeV. Then, the distribution was represented by a function composed of a gaussian and a log-normal components to describe the quiescent and flaring states, respectively. This function is shown in figure \ref{fig:Tluczykont} with fluxes above an energy threshold of 400 GeV instead of 1 TeV for comparison with the classification of VERITAS flux states presented by \citet{2014APh....54....1A}. For the conversion we assume the VERITAS spectra for the Crab and Mrk 421 \citep{2006APh....25..391H,2011ApJ...738...25A}. Although the distribution is derived from an inhomogeneous data set biased toward bright flares that induces systematic errors hard to quantify, it shows important features to point out. 
First, the upper limit on the quiescent flux of Mrk 421 reported  by \citet{Tluczykont2010} as the mean of the gaussian ($\approx$ 0.73 Crab above 400 GeV) is comparable to the VERITAS ``very low'' state, meaning that observed fluxes below the VERITAS ``very low'' state are dominated by emission of the source in the quiescent state.
Second, fluxes higher than the VERITAS ``low" state correspond to flaring states since they are described by the log-normal function. 
Third, the highest observed value for a daily-wise flux is $\approx$8 Crab for energies above 400 GeV \citep{Aharonian2003}. Because of limitations in the available energy to maintain or increase the observed VHE $\gamma$-ray flux, this value is probably significantly higher and slightly smaller than the expected at larger (weeks, months, etc) and shorter (hours, minutes, etc) time scales, respectively. Thus, a representative set of observed fluxes of Mrk 421 activity should sample fluxes as low as tenths of Crab to values close to at least 8 Crab (for energies above 400 GeV). And finally, fluxes above the VERITAS ``high" \footnote{We will refer as the VERITAS ``high" state the value given by  \citet{2011ApJ...738...25A} as a ``high state A" which corresponds to the lowest of the high states reported.} state are already between the highest fluxes, and they are at least one order of magnitude less common fluxes when compared with the VERITAS ``low" flux. Notice that these flux values do not have physical meaning but instead serve as reference for the flux magnitudes observed for Mrk 421. 

Considering the mentioned features of the flux distribution, we focus our analysis in three data sets that allow us to compare correlations obtained with different experiments and on different time scales. In particular, we use data from two already published correlations, i.e. the monthly Whipple-RXTE/ASM correlation \citep{2014APh....54....1A} with the most unbiased and comprehensive VHE $\gamma$-rays data set and the hourly HEGRA CT1-RXTE/ASM correlation \citep{Aharonian2003} with a very good flux sampling over a large range of flux values of one of the brightest flares observed.
Furthermore, we add data taken with the Milagro observatory \citep{2014ApJ...782..110A}, combining them with archival data from RXTE/ASM, based on simultaneous VHE/X-ray observations. 
The addition of Milagro data to the study of the VHE $\gamma$-ray/X-ray correlation is important for several reasons. Although Milagro data have large uncertainties, they cover a wide range of VHE $\gamma$-ray fluxes, from $\sim$ 0 to $\sim$ 6 Crabs. Furthermore,  Milagro data are reported on $\sim$ weekly time scale, i.e. on an intermediate time scale between months (Whipple) and hours (HEGRA-CT1): these features allow us to understand if differences between the Whipple-RXTE/ASM and HEGRA CT1-RTE/ASM correlations could be attributed to the different time scales considered. Finally, thanks to the long term continuous monitoring of Mrk 421, the available Milagro data can be combined to study the correlation also on $\sim$ monthly time scale: this allows to further test the robustness of the monthly correlation reported by Whipple and to eventually verify if it is intrinsic of the source or due to, i.e., systematic effects.

It is important to point out that we base our analysis on X-ray data collected with RXTE/ASM (between 2-10 keV) to avoid other sources of systematics not only because of the differences of the instrument responses at different energy ranges but also because of the spectra modeling of Mrk 421 along these energy ranges. However, later in this section, as a test of our results, we include previous studies with XMM-Newton data when a lack of VHE $\gamma$-ray/X-ray correlation is reported. We have chosen to present RXTE/ASM data in cts/s for a direct comparison with the results shown by \cite{2014APh....54....1A}. The Crab nebula flux is about 75 cts/s as given in the RXTE/ASM database\footnote{http://xte.mit.edu/asmlc/}.

\subsection{The Whipple-RXTE/ASM data.}  

\cite{2014APh....54....1A} reported on a 14-year monitoring of Mrk 421 with the Whipple 10 m telescope and combined the Whipple data with X-ray data taken with RXTE/ASM. They found that the VHE $\gamma$-ray fluxes are linearly correlated with the X-ray fluxes on monthly and yearly time scales. However, it is interesting to note that the correlation exhibits a considerable scatter: several data points (with small uncertainties and the highest $\gamma$-ray flux value) present large deviations from the best fit line and its dispersion seems to be greater for higher fluxes. 
In Sec. \ref{sec:corr} we study the monthly correlation to estimate how robust it is and how significant the deviations are from it.

\subsection{The HEGRA CT1-RXTE/ASM data.}

\citet{Aharonian2003} reported on the monitoring of Mrk 421 with HEGRA CT1 on 2001, during a period of strong activity of the source and combined the HEGRA CT1 data with X-ray data taken with RXTE/ASM. They found that the VHE $\gamma$-ray fluxes are linearly correlated with the X-ray fluxes on hourly time scale. This data set allows to test the correlation up to very high VHE $\gamma$-ray fluxes, since the 2001 flare of Mrk 421 is one of the most intense observed.

\subsection{The Milagro-RXTE/ASM data.}

\citet{2014ApJ...782..110A} showed the Mrk 421 light curve at TeV energies on $\sim$ weekly time scale obtained with the Milagro experiment, that continuously monitored the source in the period between 2005 September 21 and 2008 March 15. Here we study the correlation between these Milagro data and X-ray data taken with RXTE/ASM in the same observation period.  To obtain the VHE $\gamma$-ray/X-ray correlation we combine Milagro data with archival data from the RXTE/ASM database. In this database, each data point represents the one-day average flux of the source and is quoted as nominal 2-10 keV rate in ASM counts per second (the Crab nebula flux is about 75 ASM counts/s). We select data collected during the same observation period as Milagro, then we calculate the average X-ray rate for each of the $\sim$ weekly time bin of Milagro. Finally, we calculate the average Milagro and RXTE/ASM fluxes for consecutive groups of 4 $\sim$ weekly time bins, to study the correlation on $\sim$ monthly time scale.

\subsection{Other data sets.}

 As a test of our results, we take other available data sets reported in \citet{2009ApJ...703..169A} and \citet{2007ApJ...663..125A}.  On one hand, the data sets reported by \citet{2009ApJ...703..169A} have been used to study the VHE $\gamma$-ray/X-ray correlation on a time scale of $\sim$ 20-30 minutes during two flaring states: one on April 2006, with VHE $\gamma$-ray data from MAGIC and Whipple and, the other on May 2008 with data from VERITAS. In both periods the X-ray data correspond to XMM-Newton observations in the energy range of 0.1-20 keV, similar to RXTE/ASM data. In particular, the first period is interesting to our study since \citet{2009ApJ...703..169A} pointed out a lack of correlation among these data with a Pearson correlation coefficient of $-0.05$.  On the other hand, \citet{2007ApJ...663..125A} found VHE $\gamma$-ray and X-ray data from MAGIC and RXTE/ASM observations from November 2004 to April 2005 to be correlated on nightly time scale with a coefficient of 0.64 \citep{2007ApJ...663..125A}. 

We convert the XMM-Newton rates in RXTE/ASM rates with the online WebPIMMS\footnote{http://heasarc.gsfc.nasa.gov/cgi-bin/Tools/w3pimms/w3pimms.pl} tool, using the X-ray spectra and the hydrogen column density reported in \citet{2009ApJ...703..169A}. 

\par\medskip

The published VHE $\gamma$-ray fluxes from the different experiments have different energy thresholds and are in different units: the Whipple data are fluxes above 400 GeV in Crab units, while the Milagro and HEGRA CT1 data are fluxes above 1 TeV in cm$^{-2}$ s$^{-1}$.
Considering that the spectrum of Mrk 421 can be described by the same function (power law with a cut off around 5 TeV) from hundreds of GeV to a few TeVs, to compare all the correlations, the measured VHE $\gamma$-ray fluxes is converted to a common energy threshold and expressed in the same units. In this paper we choose an energy threshold of 400 GeV and the Crab units, as done with the Whipple data. 

The factor to convert the fluxes of a source from an energy threshold $E_0$ to another energy threshold $E_1$ in Crab units is given by 
\begin{eqnarray}
F(E > E_1)=  & \\ \nonumber  &F(E > E_0)\times \frac{\int_{E_1}^{\infty} \phi(E) dE} {\int_{E_0}^{\infty}\phi(E) dE \times \int_{E_1}^{\infty}\phi_{\rm Crab}(E) dE},
\end{eqnarray}
where $\phi(E)$ and $\phi_{\rm Crab}(E)$ are the source and the Crab spectrum, respectively. 

We take the spectra observed by the corresponding instrument and when possible for the same time period of the corresponding observation to minimize discrepancies between different data sets due to systematic effects of the experiment. If the Crab spectrum were constant in time, the observed Crab fluxes used to normalize Mrk 421 fluxes would be very similar along time and between instruments. The highest and lowest Crab fluxes above an energy of 400 GeV used differ by $\approx 22\%$ and correspond to the data with the largest uncertainties, Milagro and Hegra CT1 data respectively, reflecting the limited detector response. Otherwise, the Crab flux above an energy of 400 GeV, used in this work, differs less than $8\%$. This includes data from MAGIC, VERITAS and Whipple presented in Sec. \ref{sec:corr}. In particular, we use the Mrk 421 spectra measured by Milagro and HEGRA CT1 in the same observation period considered for the correlations (see \citealp{Aharonian2003,2014ApJ...782..110A}). We also use the average Milagro and HEGRA CT1 Crab spectra reported in \citet{2012ApJ...750...63A} and \citet{hegracrab} respectively. 
For data sets reported in \citet{2009ApJ...703..169A} and \citet{2007ApJ...663..125A} we use the Mrk 421 spectra reported in \citet{2009ApJ...703..169A} and \citet{2007ApJ...663..125A} and the Crab spectra reported in \citet{2006APh....25..391H,2008ICRC....2..691G} and \citet{2008ApJ...674.1037A}. 

\section{The VHE $\gamma-$ray/X-ray correlation}\label{sec:corr}

This section describes an alternative analysis of the Whipple-RXTE/ASM correlation, the study of the Milagro-RXTE/ASM correlation and the comparison among these correlations and the one obtained with HEGRA CT1-RXTE/ASM data. 

We use the maximum likelihood approach to estimate how robust is the monthly Whipple-RXTE/ASM correlation and how significant the deviations are from it. We consider that the correlation can be affected by an intrinsic scatter $\sigma_A$ of unknown nature that has to be taken into account. Therefore, we use the maximum likelihood method as discussed by \citet{2005physics..11182D}.  This method assumes that, if the correlated data ($F_{x,i}$, $F_{\gamma,i}$) can be described by a linear function $F_{\gamma}=a F_{x}+b$ with an intrinsic scatter $\sigma_A$, the optimal values of the parameters ($a$, $b$ and $\sigma_A$) can be determined by minimizing the minus-log-likelihood function, in which the uncertainties on $F_{x,i}$ and $F_{\gamma,i}$ ($\sigma_{F_{x,i}}$ and $\sigma_{F_{\gamma,i}}$ respectively) are taken into account. This function can be written as
\begin{eqnarray}
L(a,b,\sigma_A) = & \\ \nonumber  &\hspace{-2cm} \frac{1}{2} \sum_i \log(\sigma_A^2+\sigma_{F_{\gamma,i}}^2+a^2 \sigma_{F_{x,i}}^2)+ \frac{1}{2} \sum_i \frac{(F_{\gamma,i}-a F_{x,i}-b)^2}{\sigma_A^2+\sigma_{F_{\gamma,i}}^2+a^2 \sigma_{F_{x,i}}^2}.
\end{eqnarray}

\begin{figure*}
\begin{center}
\includegraphics[width=18cm, height=14cm]{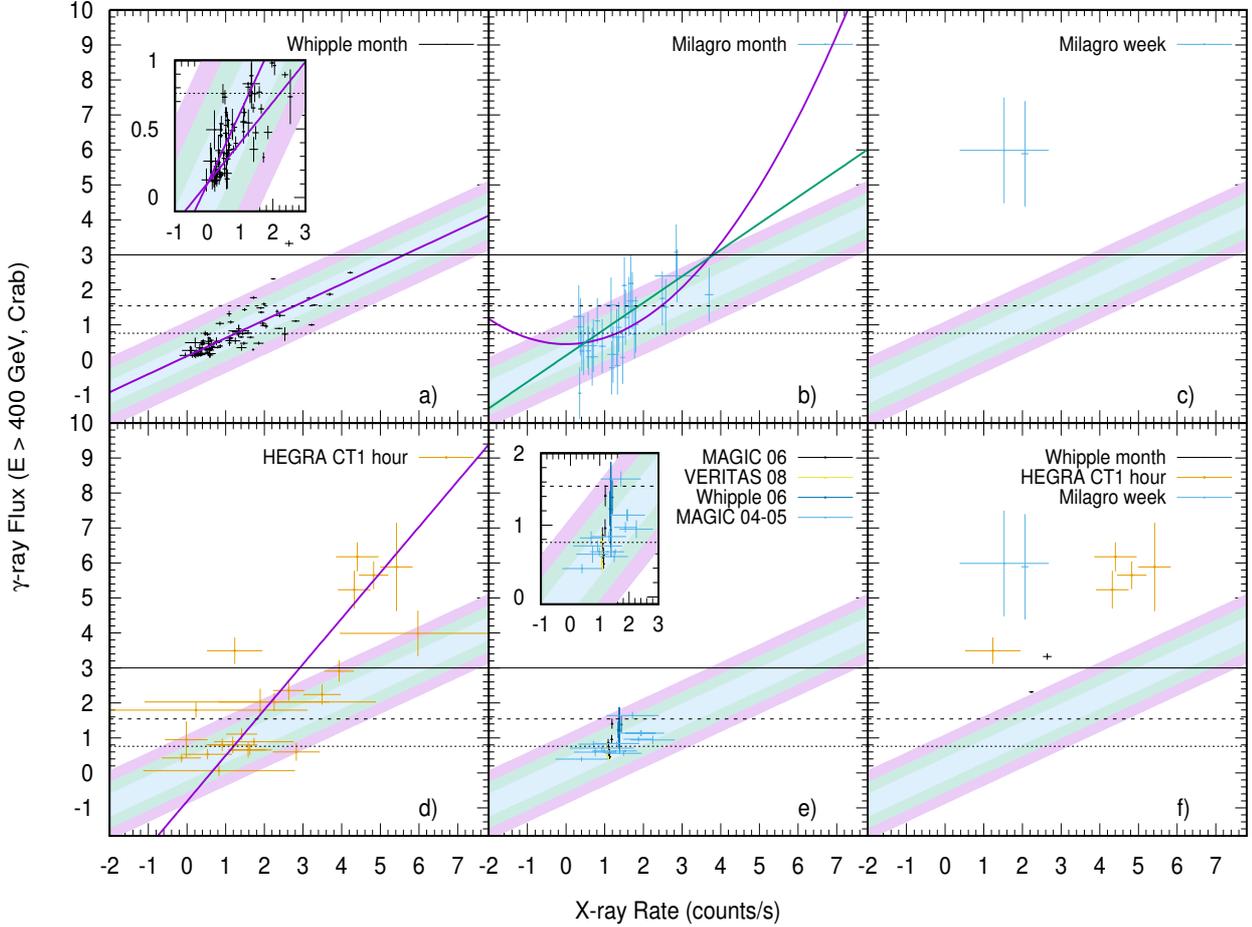}
\caption{Panel a): Correlation between the X-ray (2-10 keV, RXTE/ASM) and the VHE (E $>$ 400 GeV, Whipple) fluxes of Mrk 421 on monthly time scale, with data from \citet{2014APh....54....1A}. The purple solid line is the best fitting straight line. As for the other panels, the shadowed regions represent the 1, 2 and 3 $\sigma_A$ scatter around the best fit of this correlation; the ``very low'', ``low'' and ``high'' state flux levels of Mrk 421, as measured by VERITAS \citep{2011ApJ...738...25A} are shown as black dotted, dot-dashed and solid lines respectively. A zoom in the region of the lower X-ray and VHE fluxes is also shown, with the red dot-dashed line representing the best fitting straight line of the points with VHE $\gamma$-ray flux up to 1 Crab. Panel b): Correlation between the X-ray (2-10 keV, RXTE/ASM) and the VHE (E $>$ 400 GeV, Milagro) fluxes of Mrk 421 on $\sim$ monthly time scale, with data from \citet{2014ApJ...782..110A} and the RXTE/ASM public database. The purple solid line represents the best fitting straight line (see text), while the green line is the best fitting quadratic function (see text). Panel c) Correlation between the X-ray (2-10 keV, RXTE/ASM) and the VHE (E $>$ 400 GeV, Milagro) fluxes of Mrk 421 on $\sim$ weekly time scale, with data from \citet{2014ApJ...782..110A} and the RXTE/ASM public database. Panel d): Correlation between the X-ray (2-10 keV, RXTE/ASM) and the VHE (E $>$ 400 GeV, HEGRA CT1) fluxes of Mrk 421 on hourly time scale and its best fitting straight line (in purple) as reported in \citep{Aharonian2003}, with data from \citet{Aharonian2003}. Panel e): Correlations between RXTE/ASM rates and MAGIC (in grey), Whipple (in black) and VERITAS (in cyan) fluxes, with data from \citet{2009ApJ...703..169A}, together with MAGIC-RXTE/ASM correlation reported in \citet{2007ApJ...663..125A} (in orange). A zoom in the region of the lower X-ray and VHE fluxes is also shown. Panel f): Points of the monthly Whipple-RXTE/ASM, weekly MILAGRO-RXTE/ASM and HEGRA CT1-RXTE/ASM data sets lying above more than 3 $\sigma_A$ from the best fit straight line of the Whipple-RXTE/ASM correlation, together with the Whipple-RXTE/ASM daily point corresponding to one of the strongest flare observed by Whipple (see text).}\label{fig:whipple}
\end{center}
\end{figure*}

With this method we obtain that the data are positive correlated within 3 $\sigma_A$, with a Pearson correlation factor of $0.8$ and p-value $< 1\times10^{-16}$ and described by $a= 0.51 \pm 0.04$, $b=0.11 \pm 0.06$ and $\sigma_A=0.33 \pm 0.03$. 
The results are shown in Fig. \ref{fig:whipple}a.
We observe that there are two intense VHE $\gamma$-ray fluxes that seem not to be accompanied by intense X-ray fluxes; they lie at 4 and 6 $\sigma_A$ above the best linear fit. A similar feature has been mentioned by \citet{2003ApJ...598..242A}. They found a positive correlation between data from Tibet Air Shower Array and RXTE/ASM on monthly time scale, but noticed the presence of some intense TeV fluxes that seem to have occurred without intense X-ray fluxes falling out of their found correlation. The significance of the deviation of these observations were not quantified by \citet{2003ApJ...598..242A} and direct comparison between the Tibet-RXTE/ASM correlation and the Whipple-RXTE/ASM correlation cannot be performed because of the unknown conversion factor between the two sets of VHE $\gamma$-ray data. Deviations for the correlation are shown for other data sets later in this section and discussed widely in Section \ref{sec:break}.

It is observed from the insert in Fig. \ref{fig:whipple}a that the data dispersion around the best fit straight line is smaller in the region of low VHE $\gamma$-ray fluxes (below the VERITAS ``very low'' state flux), where almost all the points lie within 1 $\sigma_A$. This is surprising considering that the measurement of such low fluxes should be dominated by the detector sensitivity threshold effect, i.e. lower fluxes are expected to present larger uncertainties and larger dispersion. Moreover, the best fit considering only VHE $\gamma$-ray fluxes below 1 Crab (dot-dashed line in the insert of figure \ref{fig:whipple}a) is consistent within 3 $\sigma_A$ with the general correlation up to X-ray fluxes of $\sim$ 4.5 counts/s. 
\cite{Aleksic2015} has reported a positive correlation with not time lag between Whipple VHE $\gamma$-ray fluxes as low as half of the Crab Nebula and Swift data from the multi-wavelength campaign between January 2009 and June 2009. Our result confirms the existence of the correlation at low fluxes or quiescent states, as reported by Magic, but we also find that it seems to be the same as for flaring states.

In order to further understand the X-ray/VHE $\gamma$-ray monthly correlation we analyze Milagro monthly data combined with simultaneous RXTE/ASM data as done for Whipple data. A positive correlation, weaker than the one obtained with Whipple data but significant, is observed between the VHE $\gamma$-ray and the X-ray fluxes, with a Pearson correlation coefficient $R=0.67$ and p-value of $2.1\times10^{-5}$ (see Fig. \ref{fig:whipple}b). Using the maximum likelihood approach we obtain F$_\gamma$=(0.76 $\pm$ 0.06) F$_x$ - (0.12$\pm$ 0.10), with $\sigma_{A,M}$=0.0001 (meaning that there is no need to consider an extra scatter to find a consistent fit with the data), while a quadratic fit yields  F$_\gamma$=(0.19$\pm$ 0.04) F$_x^2$ + (0.45 $\pm$ 0.16), with $\chi^2$/$d.o.f.$=22.4/28 ($d.o.f.$ is the number of degrees of freedom). As observed, a more complex description as the quadratic function is not required by the data. We compare this correlation with the one obtained with Whipple data, finding that they are consistent within 3 $\sigma_A$ (see Fig. \ref{fig:whipple} b). In other words, the monthly correlation observed by Whipple is supported by Milagro data and it is the same from TeV fluxes of tenths of Crab up to flaring state fluxes of 2-3 Crabs.

Correlation on fluxes measured on monthly and shorter time scales may not be necessarily the same. In fact, fluxes averaged over longer time scales  (i.e. months)  are dominated predominantly by the quiescent or lowest states, while fluxes averaged over shorter time scales (i.e. weeks to hours) are measured in flaring states because of the instrument sensitivities. It may be that different physical processes are responsible for the emission when the source is in different levels of activity, possibly leading to different correlations. Thus,
we use Milagro and HEGRA CT1 data to test the correlation found at shorter time scales. First, we obtain simultaneous RXTE/ASM and Milagro data on $\sim$ weekly time scale and we compare them with the monthly Whipple-RXTE/ASM correlation.  Due to the large uncertainties of the Milagro data, in this case we only consider the VHE $\gamma$-ray fluxes with a significance above the background greater than 3 (before trial corrections). It comes out that there are only two points with this characteristic and they lie at about 10 $\sigma_A$ from the best fitting straight line of the Whipple-RXTE/ASM correlation (see Fig. \ref{fig:whipple}c). Surprisingly, these Milagro-RXTE/ASM data are associated to the flares occurred in 2006 (see further discussion in section \ref{sec:break}) and also point towards intense VHE $\gamma$-ray fluxes without intense X-ray fluxes as the Whipple data deviating from the correlation. 

Now, we do a comparison with the HEGRA CT1-RXTE/ASM correlation on hourly time scale (see  Fig. \ref{fig:whipple}d) as reported by \citep{Aharonian2003}. It can be seen that considering only the HEGRA data for the gamma ray emission, the correlation looks steeper\footnote{the index of the best fitting power-law function reported by \citet{Aharonian2003} is $a$=1.98$\pm$0.18, that corresponds to 1.31$\pm$0.12 when HEGRA CT1 data are converted into fluxes above 400 GeV in Crab units.} than when considering only Whipple data. However, it is observed that most of the data is consistent within 3 $\sigma_A$ with the monthly correlation. Surprisingly again, there are 5 data points, corresponding to the highest TeV fluxes, also higher than the VERITAS ``high state'' flux,  that lie above more than 3 $\sigma_A$ from the best fit of the Whipple-RXTE/ASM correlation, as also observed with the weekly Milagro-RXTE/ASM data. 

 As a test to the correlation (that seems to  stand for any time scale), we take other available data sets reported in \citet{2009ApJ...703..169A} and \citet{2007ApJ...663..125A} and shown in figure \ref{fig:whipple}e.  Consistently with our previous results, all TeV fluxes below the VERITAS ``low state'' flux lie within 1 $\sigma_A$ from the found Whipple-RXTE/ASM correlation and the scatter in the data increases as the TeV fluxes increase. However, it can be seen that all data lie within 3 $\sigma_A$ from the correlation and that the lack of correlation reported before by  \citet{2009ApJ...703..169A} can be understood, now, as the result of the small coverage of the TeV fluxes range observed. We do not find data points deviating from the correlation, but there are no TeV fluxes as high as the VERITAS ``high state'' either, a common characteristic in the previously discussed Whipple, Milagro and Hegra CT1 data sets, see figure \ref{fig:whipple}f.

The results here discussed are based on the comparison of various data sets with the Whipple-RXTE/ASM correlation, taken as a reference since, as explained in Sec. \ref{sec:datasets}, it has been obtained with the most unbiased and comprehensive VHE $\gamma$-rays data set in the RXTE era. However, for completeness we also present the overall correlation, fitting all the data sets here presented together, with the maximum likelihood approach described at the beginning of this section.
We obtain $a=0.67 \pm 0.04$, $b= -0.01 \pm 0.05$ and $\sigma_A=0.37 \pm 0.03$, i.e. the correlation is consistent, within 3 $\sigma_A$, with the Whipple-RXTE/ASM correlation even though it presents a  higher slope and intrinsic scatter. More important is the fact that  7 of the previous 10 reported points (shown in Fig. \ref{fig:whipple}f) still lay at more than 3 $\sigma_A$ from the correlation (with a minimum deviation of $\sim$ 5 $\sigma_A$ and a maximum deviation of $\sim$ 9 $\sigma_A$). It is important to point out that a flux dependency on the spectral index, for instance a harder-when-brighter behavior, will modify the values of the fit parameters.

\begin{figure*}
\begin{center}
\includegraphics[scale=1.5]{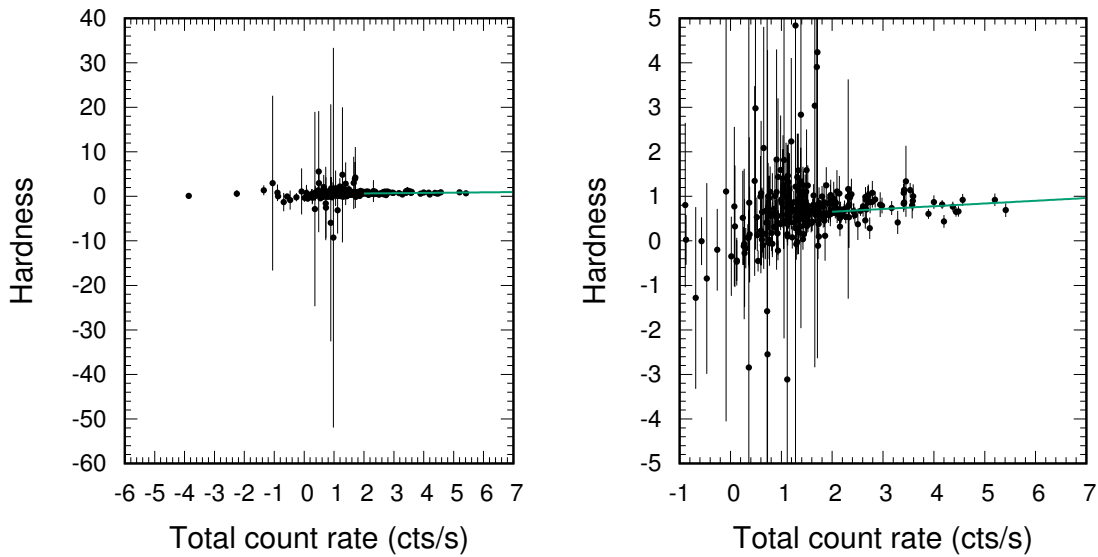}
\caption{The X-ray hardness as a function of RXTE/ASM count rate is shown. The hardness is calculated using the energy channels A (1.3 - 3.0 keV) and B (3.0 - 5.0 keV) from the ASM database and can be used to obtain a rough estimation of the spectral index \citep{2002ApJ...578..357Z}. The right-hand panel shows a zoom of the left-hand panel. A detailed description of the hardness dependency on the total count rates requires a deep and detailed analysis. However, for guidance a linear fit (green line) for total count rates above 2 cts/s is shown. A weak harder-when-brighter behavior can be observed without displaying a strong spectrum variability for total count rates above 2 cts/s.}\label{fig:spec_var}
\end{center}
\end{figure*}

We have found that the emission of Mrk 421 at energies above 400 GeV is correlated with its emission at X-rays energies between 2-10 keV within 3$\sigma_A$ at least up to $\gamma$-fluxes as high as 3 Crabs (at energies above 400 GeV). This correlation is positive and can be described by only one linear function even for the lowest fluxes observed, independently of the instrument, time scale (at least larger than hours) or period of observation. The fact that the same correlation stands independently of the time scale could be a consequence that the same correlation stands for quiescent or low flux states and flaring states (at least when the flux is below  $\rm \sim 2.5\times 10^{-10} cm^{-2}s^{-1}$). However, the scatter of the data at medium-high fluxes, included in the estimation of $\sigma_A$, has to be further studied. For instance, it could be the result of:  1. small differences among flares on the physical parameters such as magnetic field strength and injected electron density that result in deviations from the linear description; 2. uncertainties in flux measurements, mainly because they were not completely simultaneous, but the lack of counterpart measurements of high X-ray fluxes associated with low $\gamma$-ray fluxes would have to be explained; 3. high flux variability in only one of the energy bands studied here; 4. a sudden and simultaneous variability in both, $\gamma$ and X-ray spectra mainly for high fluxes, although the hardness does not show evidence of strong spectral variability for count rates above 2 counts per second as shown in Figure \ref{fig:spec_var}; 5. or due to whatever is the physical process that ends into the break of the correlation at the highest fluxes. It is important to point out that the correlation found in this study using X-rays observations at energies between 2-10 keV may or may not be valid for other, shifted or extended, X-rays energy range, see \citet{2009ApJ...695..596H}. However, the fact that the correlation stands for multiple observations in different flaring states constrains the  parameters of the one-zone SSC model as shown in Sec. \ref{sec:theo}. 

\section{break-down at the highest TeV fluxes}\label{sec:break}

In Sec. \ref{sec:corr} we have shown that monthly correlations obtained with different VHE experiments (Whipple and Milagro) and in overlapping observation periods (the 14-year observation period of Whipple includes the 3-year monitoring period of Milagro) are consistent within 3 $\sigma_A$: this excludes possible systematic effects and suggests a ``unique'' monthly correlation with information about intrinsic properties of the source and characterizes its average behavior. 
However, we have also found evidence of two points lying at more than 3 $\sigma_A$ from the best fit straight line. These two points correspond to observations made by Whipple on the first months (from February to April) of  2001, period over which Mrk 421 showed an exceptionally strong and long (months) lasting flaring activity \citep[see e.g.,][]{Aharonian2002, 2002ApJ...575L...9K}. 
Thus, the main contribution to these two monthly fluxes comes from the emission in flaring states and probably the break-down of the correlation occurs on very high levels of activity of the source. This result is supported by the comparison of the monthly data with data on shorter time scales associated to an intense activity of Mrk 421. For instance, we have shown that the correlation obtained with VHE $\gamma$-ray data taken by HEGRA CT1 during the intense flare of 2001 on time scale of hours \citep{Aharonian2003} is consistent with the monthly correlation when the low VHE $\gamma$-ray fluxes are considered; on the contrary, there are several intense VHE $\gamma$-ray fluxes not accompanied by intense X-ray fluxes, lying at more than 3 $\sigma_A$ from the best fit line of the monthly data. 

The break-down of the correlation is not only a property of the intense flare of 2001, but is observed also in other time periods and by other experiments. For instance, a similar break-down of the monthly correlation is observed with weekly Milagro-RXTE/ASM data, as shown in Fig. \ref{fig:whipple}c. These weekly points are related to data taken i) between the end of May/beginning of June 2006 and ii) the second week of August 2006. These two weeks  correspond to a period of very high activity of Mrk 421 at X-rays (see \citealp{2011ApJ...738...25A}). The observed X-ray fluxes are close to values where the correlation breaks down ($>$ 4 counts/s) and are possibly associated to an increment of the activity at VHE $\gamma$-rays. In fact, in these weeks Milagro observed the most intense fluxes of Mrk 421, although the significance is not high enough to claim for a flare detection \citep{2014ApJ...782..110A}. Other observations of VHE experiments in the overall two weeks on same or shorter time scale are not available to confirm a state of high activity of Mrk 421, but it is observed some tendency of the VHE $\gamma$-ray fluxes measured by Whipple to increase at the end of May 2006 \citep{2011ApJ...738...25A}.

Finally, \citet{2011ApJ...738...25A} reported on the observation of a possible ``orphan'' TeV flare of Mrk 421 on May 2-3, 2008. Also, \cite{2014APh....54....1A} reported on the observation, on May 2, 2008, of one of the strongest flare detected by Whipple during the whole 14-year observation period. This flare had a flux exceeding 10 and 5 Crab in 5-minutes and 1-hour time intervals, respectively.  We combine the average daily Whipple flux with the average daily RXTE/ASM flux measured the same day and show them in figure \ref{fig:whipple}f. It can be seen that also this flare lies above more than 3 $\sigma_A$ from the Whipple-RXTE/ASM monthly correlation.

Summarizing, the points with the highest VHE $\gamma$-ray fluxes, independently of the observation period and on the instrument, do not lie on the Whipple-RXTE/ASM correlation (see Fig. \ref{fig:whipple}f): this suggests that the correlation only holds up to a certain value of the TeV flux. These points do not even appear to be correlated among themselves: similar TeV fluxes are associated with very different X-ray fluxes, ranging from $\sim$ 1 counts/s to $\sim$ 5 counts/s. The identification of the exact value of the TeV flux at which the correlation breaks down require more detailed simultaneous and unbiased VHE $\gamma$-ray/X-ray observations over a wide and continuous range of TeV fluxes and over short time scales (i.e. minutes) because, observations on longer time scales can be used only to put lower limits on this TeV flux. A point lying outside the correlation (such as, i.e., for the highest monthly Whipple flux) could be the combination of the emission in a low flux state (eventually lying on the correlation if considered alone) and the emission in a high flux state (lying outside the correlation); the resulting measured TeV flux will then be lower than the one breaking down the correlation.
All the points lying outside the correlation discussed in this section have a VHE $\gamma$-ray flux greater than the flux corresponding to the VERITAS ``high state'' (3 Crab), which can then be considered as a lower limit for the break TeV flux.

The observed break-down of the correlation for the brightest VHE $\gamma$-ray flares is consistent with the results presented by \citet{2014ApJ...782..110A}. They estimated the TeV duty cycle\footnote{The duty cycle is the fraction of time that the source spends in flaring states.} ($DC$) of Mrk 421 with data from several VHE experiments (including Milagro, Whipple and HEGRA CT1)  and compared it with the X-ray $DC$, as determined by \citet{Resconi2009} with RXTE/ASM data. In general, a TeV $DC$ approximately equal to the X-ray $DC$ is expected if SSC is the physical process responsible for the VHE $\gamma$-ray emission. \citet{2014ApJ...782..110A} found that the TeV $DC$ and the X-ray $DC$ are consistent when considering the less intense flares, while the TeV $DC$ becomes greater than the X-ray $DC$ when looking at bright flares. Thus, that there are intense TeV flares not accompanied by intense X-ray flares explaining the observed break-down of the VHE $\gamma$-ray/X-ray correlation.

It is interesting to note that a similar break-down of the correlation for the VHE $\gamma$-ray fluxes has been observed also in another blazar: 1ES 1959+650. \citet{2004ApJ...601..151K, 2014ApJ...797...89A} studied the VHE $\gamma$-ray/X-ray correlation with data taken with HEGRA, Whipple and RXTE/PCU during a 3-months campaign on 2002. They shown that, although in general the X-ray and the VHE $\gamma$-ray fluxes seem to be correlated, there is a clear break-down of the correlation for the highest VHE $\gamma$-ray fluxes (flux $\rm \gtrsim$ 3 Crab at energies above 600 GeV), corresponding to what they called a ``TeV orphan flare''. 
However, it is also important to mention that an opposite behavior, i.e. a break-down of the correlation at low VHE $\gamma$-ray fluxes, has been observed as well: this is the case of the blazar PKS 2155-304. \citet{Abramowski2012} performed a 2-weeks multi wavelength campaign of this source on summer 2006 and found that the X-ray and VHE $\gamma$-ray emission are correlated during the  high state of the source, while there is no clear correlation for the nights with the lowest flux levels: again this conclusion could not stand if the flux coverage is not large enough.

\section{Theoretical model}\label{sec:theo}

TeV $\gamma$-ray and X-ray correlation can be interpreted in the SSC framework \citep{Aleksic2015}.  In this model, the emitting region moving at ultra-relativistic velocities in a collimated jet has an electron distribution. These  Fermi-accelerated electrons injected in the emitting region and confined to it by a magnetic field, radiate photons by synchrotron emission and after that scatter them up to higher energies by inverse Compton.  In order to introduce the relevant equations, the natural units will be shown ($c=\hbar=1$).   In addition,  primes and unprimed quantities are given in the comoving and observer frame,  respectively.

\subsection{Synchrotron radiation}

Fermi-accelerated electrons in the emitting region are usually described by a broken power-law written as \citep{1994hea2.book.....L}

\begin{equation}
\label{espele}
N_e(\gamma_e)   \propto
\begin{cases}
\gamma_e^{-p_1} 						& 	\gamma_e < \gamma_{\rm e,c},\cr
   \gamma_e^{-p_2}          & 	\gamma_{\rm e,c} \leq  \gamma_e<\gamma_{\rm e,max},\cr
\end{cases}
\end{equation}
\noindent where $p_1$ and $p_2$ are
the electron spectral indexes and $\gamma_{\rm e,c}$ and $\gamma_{\rm e,max}$  are cooling and maximum electron Lorentz  factors, respectively.

Electrons immerse in a magnetic field density $U'_B=\frac{B'^2}{8\pi}$ cool down in accordance with the cooling time scale $t'_c=\frac{3m_e}{4\sigma_T}\,\,{U'}^{-1}_B\,\gamma^{-1}_e$ \citep[e.g., see][]{2014MNRAS.437.2187F,2014ApJ...783...44F}.   By comparing the previous time scale  with the dynamic scale $t'_d\simeq r_d/\Gamma$ \citep{2009ApJ...704...38S,1985MNRAS.216..241B, 2018arXiv181101108F,2014MNRAS.441.1209F}, the break photon energies is given by

\noindent 
\begin{eqnarray}\label{synrad}
\epsilon^{\rm syn}_{\rm \gamma,c} &=&\frac{ 9\sqrt{2\pi}\,q_e\,m_e}{8\,\sigma_T^2}\,(1+z)^{-1}\,(1+Y)^{-2}\,\Gamma^3\, {U'}_B^{-3/2}\, r_d^{-2}\,,
\end{eqnarray}

where  $m_e$ is the electron mass, $q_e$ is the elementary charge, $\sigma_T$ is the Thomson cross section, $z=0.03$ is the redshift, $\Gamma$ is the bulk Lorentz factor, $r_d$ is the size of the emitting region and  $Y=U'_e/U'_B$ is the Compton parameter with $U'_e$ the energy density given to accelerate electrons.   The observed synchrotron spectrum is calculated explicity in \cite{2016ApJ...830...81F}.

The values of cosmological parameters adopted are given in  \cite{2018arXiv180706209P}.

\subsection{Inverse Compton scattering}

Fermi-accelerated electrons in the emitting region can up-scatter synchrotron photons up to higher energies. From  eq. (\ref{synrad}), we obtain that the break energies of the SSC emission are

\begin{eqnarray}\label{icrad}
\epsilon^{\rm ssc}_{\rm \gamma,c} &=&\frac{ 81\sqrt{2\pi}\,q_e\,m_e^3}{128\,\sigma_T^4}\,(1+z)^{-1}\,(1+Y)^{-4}\,\Gamma^5\, {U'}_B^{-7/2}\, r_d^{-4}.
\end{eqnarray}

The inverse Compton scattering spectrum from synchrotron radiation is explicitly calculated in \cite{2016ApJ...830...81F}. 
The effect of the extragalactic background   light (EBL) absorption modeled by \cite{2008A&A...487..837F} is included.

\subsection{X-ray / TeV \texorpdfstring{$\gamma$}-ray  Correlation}

 In order to obtain the parameter values (a, b and $\sigma_A$) that reproduce the VHE $\gamma$-ray / X-ray lineal correlation (see Section \ref{sec:corr}), the power law $\epsilon^{\rm syn}_{\gamma,c} < \epsilon_\gamma$ of the synchrotron spectrum is used to describe the X-ray flux and the power law $\epsilon^{\rm ssc}_{\rm \gamma,c} < \epsilon_\gamma$  of the inverse Compton spectrum to describe VHE $\gamma$-ray flux.  In addition,  the  units of synchrotron spectrum are converted from erg/cm$^2$/s to counts/s using the online WebPIMMS \footnote{http://heasarc.gsfc.nasa.gov/cgi-bin/Tools/w3pimms/w3pimms.pl} tool for  a hydrogen column density of 1.61 $\times$ 10$^{20}$ cm$^{-2}$\footnote{This is an intermediate value between the Leiden/Argentie/Bonn (LAB) Survey and Dickey and Lockman (DL) value as given in the WebPIMMS tool.}, and the units of inverse Compton spectrum are converted to Crab units with the Crab spectrum as measured by Milagro \citep{2012ApJ...750...63A}.   
  For the electron spectral index, $p_2$, a value of 3.6 is assumed. The corresponding spectral index of 2.3 is within the reported ones in the TeV observations considered in this work.

\begin{figure*}
\begin{center}
\includegraphics[scale=0.6]{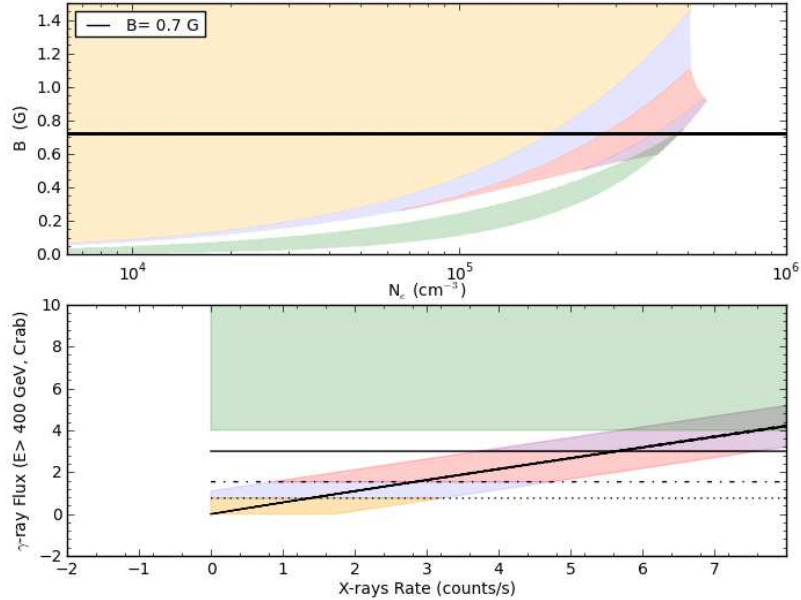}
\caption{Parameter regions of magnetic field (B) and electron density (Ne) (upper panel) for which the VHE $\gamma$-ray and X-ray fluxes are correlated (lower panel). The regions in yellow, purple and pink represent the parameter spaces that allow to reproduce the observed correlation up to the very low, low and high states, respectively. The violet and the green region represents the parameter spaces that allow to reproduce the observed correlation above the high state and the points breaking down the correlation, respectively. The ``very low'', ``low'' and ``high'' state flux level of Mrk 421 as measured by VERITAS \citep{2011ApJ...738...25A} are shown as black dotted, dot-dashed and solid lines respectively.}\label{fig:teo}
\end{center}
\end{figure*}

A bulk Lorentz factor of $\Gamma$ = 10, a emitting radius of $r_d=5.0 \times 10^{16}$ cm \citep[see e.g.,][]{2014ApJ...782..110A, 2011ApJ...734..110B,2011ApJ...736..131A} and values for the magnetic field (B) and the electron number density (Ne) in the ranges of 0.05 G $\leq B' \leq$ 2 G and 10$^3$ cm$^{-3} \leq$ Ne $\leq$ 10$^{4.5}$ cm$^{-3}$, respectively, were taken. These ranges are the ones successfully used by several authors \citep{Qian1998,Tramacere2007,2011ApJ...734..110B,2011ApJ...736..131A,2013PASJ...65..109C} to describe the broadband SED  of Mrk 421 within the SSC scenario.  The Klein-Nishina effect which is included in this model results in a  reduction of the Compton scattering cross-section. It occurs when the incoming photon is large compared to the rest mass energy of the electron. The optical depth ($\tau_{\gamma\gamma}$) to pair production by $\gamma\gamma$ interaction is proportional to the emitting region and the target photon, $\tau_{\gamma\gamma}\simeq \sigma_T\,n(x) x r_d$ \citep{1998MNRAS.293..239C}, where $x$, in electron mass units,  is the energy of target photons with number density $n(x)$. Given the previous values and the X-ray peak at ($\sim$ keV), the optical depth is less than one.  Figure \ref{fig:teo} shows the allowed range values for the magnetic field and the electron density number that yield flux values within 3$\sigma_A$ of the observed X-ray/VHE $\gamma$-ray correlation.

For discussion purposes colors indicate $\gamma$-ray flux intervals defined by the "very low", "low"and "high" state flux levels as measured by VERITAS in \citet{2011ApJ...738...25A}. It can be seen that for the region of low X-ray/VHE $\gamma$-ray fluxes (in yellow) there is a wide range of values of B and Ne that allow to reproduce fluxes consistent with the observed correlation; on the contrary, for the higher fluxes the regions of allowed parameters (in purple and pink) become smaller. In other words, if we have one of the strongest flares, as the one observed by \citet{Aharonian2003}, a narrower range of values of $B$ between $0.6$ and $0.8\,G$ reproduces the overall correlation (from low to high fluxes) even when other values of the spectral index between 2.4 and 2.7, as reported in the TeV observations, are considered. For different values of $\Gamma$ and $r_d$ the same behavior is observed, as the correlation extends to larger fluxes it is possible to restrict the values of  the magnetic field. Moreover, in a given flare, any large deviation from the values of $B$ will be reflected in the maximum $\gamma$-ray flux that such flare can reach. In fact, fluctuations within the allowed range of values of $B$ between flares could be causing part of the scatter quantified by $\sigma_A$.

Green zone in Fig. \ref{fig:teo} indicates the parameters region required to explain the outliers of the correlation discussed in Sec. \ref{sec:break}. As observed, this region does not overlap with the allowed region for the lower TeV $\gamma$-ray fluxes (yellow, blue and pink zones). In other words, within the one zone SSC model if there is a flare which fluxes lay outside of the correlation for a given period of time, a large and sudden decrement followed by a restoring increment of the magnetic field is required. Shocks travel long distances increasing the probability of finding field turbulences encouraging to think that magnetic field variations may be correlated with the inhomogeneities in the jet or in the plasma that can compress and re-order the complex structure of the magnetic field \citep{1980MNRAS.193..439L}. If that is the case, this evolution of the magnetic field structure in the emitting region would result in variations of the polarization degree and position angle \citep{1985MNRAS.216..241B} as the ones observed by \citet{1990AAS...83..183M, 1991AAS...90..161T, 1992AAS...94...37T,2017ApJS..232....7F}.

\section{Conclusions}\label{sec:concl}
In this paper we presented a robust and comprehensive study of the VHE $\gamma$-ray/X-ray correlation of the blazar Mrk 421 on different time scales: from hours to months. We used X-ray data mainly from RXTE/ASM instrument to avoid more sources of systematics. 

First we used a maximum likelihood approach to perform a detailed statistical study of the reported Whipple-RXTE/ASM monthly correlation \citep{2014APh....54....1A}, to quantify the dispersion of the data. We found that in general the data are well correlated within 3 $\sigma_A$, an extra intrinsic scatter estimated with the maximum likelihood method discussed by \citet{2005physics..11182D}. However, we also found evidence of two points lying at more than 3 $\sigma_A$ from the best fitting straight line: two intense VHE $\gamma$-ray flares not accompanied by intense X-ray flares. We then combined Milagro data \citep{2014ApJ...782..110A} with archival RXTE/ASM data and showed that they are well correlated on monthly time scale; furthermore, we found that this correlation is consistent, within 3 $\sigma_A$, with the Whipple-RXTE/ASM correlation. This strongly suggests the existence of a ``unique'' monthly correlation. Finally, we compared VHE $\gamma$-ray/X-ray correlation on different time scales. To do this, we used Milagro and RXTE/ASM data on weekly time scale and we compared them with the monthly Whipple-RXTE/ASM data and with the hourly HEGRA CT1-RXTE/ASM data  \citep{Aharonian2003}. We found that the VHE $\gamma$-ray/X-ray correlation becomes steeper when shorter time scales are considered. We compare previous results on correlations between $\gamma$- and X - ray emissions and found that after converting all $\gamma$ fluxes to Crabs and to give them above the same energy threshold, all are consistent with our results. Even those reporting a lack of correlation are explained as a short coverage of the flux ranges.

We have used the  ``standard" one zone SSC model to interpret the observed correlation and found that it can be explained as a consequence of an intrinsic property of the source such as the magnetic field, B. 
 We found that a description of a correlation that extends to larger flux values, considerately restricts the values of the magnetic field. In particular, for a $\Gamma$ of 10 and a spectral index with a constant value of 2.3 along the fluxes, $B$ between  $0.6$ and $0.8$ Gauss describes flares evolving from the lowest to the highest fluxes within the correlation. Fluctuations of the magnetic field within the mentioned range between flares is a possible cause of the scatter measured by $\sigma_A$ or/and of the steepness, if real, of the correlation at smaller time scales.  On the other hand, the break-down of the correlation is hard to explain within the "standard" one zone SSC model without large changes in the jet parameters characterizing the SSC emission and/or an extra contribution of hadronic processes to the VHE $\gamma$-ray emission; further studies are needed to fully understand their nature.

All these results represent an important tool to understand the nature of the physical processes responsible for the VHE emission of Mrk 421. However, more sensitive TeV observations are necessary to confirm these findings, to identify exactly at what time scale the break-down of the monthly correlation starts and to put tight constraints on the emission models. With the High Altitude Water Cherenkov detector (HAWC, see e.g. \citealp{2013APh....50...26A}), it will be possible to study with more accuracy the VHE $\gamma$-ray/X-ray correlation of Mrk 421: thanks to its large sensitivity it will allow to better characterize the correlation when the source is in low states of activity, as well as to compare correlations obtained in different observation periods and on different time scales.

\section*{Acknowledgements}

We thank P. Moriarty for the data of the monthly Whipple-RXTE/ASM correlation. This work has been supported by UNAM PAPIIT grants IA102917 and IG100317.






 \newcommand{\noop}[1]{}






\end{document}